\documentclass[12pt]{article}

\usepackage{amsmath}
\usepackage{amsfonts}

\newtheorem{theorem}{Theorem}

\newtheorem{definition}[theorem]{Definition}

\begin{document}

\title{Clustering by hypergraphs and \\ dimensionality of cluster systems}

\author{S.Albeverio\footnote{University of Bonn, Germany}, S.V.Kozyrev\footnote{Steklov Mathematical Institute, Russian Academy of Sciences}}

\maketitle

\begin{abstract}
In the present paper we discuss the clustering procedure in the case where instead of a single metric we have a family of metrics. In this case we can obtain a partially ordered graph of clusters which is not necessarily a tree. We discuss a structure of a hypergraph above this graph. We propose two definitions of dimension for hyperedges of this hypergraph and show that for the multidimensional $p$-adic case both dimensions are reduced to the number of $p$-adic parameters.

We discuss the application of the hypergraph clustering procedure to the construction of phylogenetic graphs in biology. In this case the dimension of a hyperedge will describe the number of sources of genetic diversity.
\end{abstract}

\section{Introduction}

The clustering procedure describes the construction of a partially ordered tree of clusters (or hierarchy) starting from a metric on a set of points \cite{Murtagh}.

In the present paper we investigate the following problem. Assume we have instead of a single metric a family of metrics depending on a set of parameters (this is a typical situation in applications). We will obtain a family of clusterings. What is the structure of this family? Can we describe this family by a single mathematical object? We discuss the approach to clustering based on an application of partially ordered hypergraphs.

We start with a pair of examples of hypergraph clustering and then propose a general definition.
Our definition is based on the following observation: for two different clusterings which correspond to the different metrics it may happen that some clusters (with respect to the different metrics) coincide as sets. This allows to unify the different clustering trees into a single partially ordered graph. Moreover it is natural to consider a structure of a hypergraph on this graph where the hyperedges will describe the alternative ways of growth of a cluster with the increase of its diameter (with respect to the different metrics).

We give a general description of this hypergraph and apply it to a discussion of multidimensional structures in data. Our motivating example is given by the family of different metrics in $\mathbb{Q}_p^d$ which combines both the hierarchy and the multidimensional structure. We propose two definitions of dimension for hyperedges of a hypergraph of clusters. Both definitions of dimensions (the A--dimension and the B--dimension, see the section 5 below) in the $p$-adic case reduce to the number of $p$-adic parameters (in particular in this case these dimensions coincide).

In data analysis trees of clusters are used for classification purposes and describe the diversity in data. One of the important applications of clustering is the application to construction of phylogenetic trees using the analysis of genomic sequences. The procedure of hypergraph clustering discussed in the present paper allows to describe the situation when we have several sources of diversity. In particular the dimension of hyperedges describes the number of the sources of diversity for the corresponding data. In bioinformatics this might be helpful in the situation where the analysis of the different parts of a genome generates different phylogenetic trees (in particular, for the discussion of a ''forest of life'' instead of a ''tree of life'' \cite{TreeLife,Koonin}). This behavior is typical for the cases of reticulate evolution (in particular, hybridization and horizontal gene transfer), where instead of phylogenetic trees one has to consider phylogenetic networks.

An example of hypergraph of clusters for clustering with respect to a pair of metrics was discussed in \cite{Hcluster}. A family of multidimensional ultrametrics on $\mathbb{Q}_p^d$ was investigated in \cite{quincunx} in relation to multidimensional $p$-adic wavelets with matrix dilations. Analysis in general locally compact ultrametric spaces and wavelets on these spaces were discussed in \cite{ACHA}. For a review of ultrametric mathematical physics see \cite{review}.

The exposition of the present paper is as follows.

In section 2 we discuss two simple examples of hypergraph clustering.

In section 3 we discuss hypergraph clustering for multidimensional $p$-adic spaces.

In section 4 we give general definitions of hypergraph of balls and of dimensions of hyperedges for a general ultrametric space with a family of ultrametrics.

In section 5 we discuss applications of hypergraph clustering and dimensions of hyperedges for phylogenetic graphs.

In section 6 (Appendix) we recall the clustering procedure and the construction of duality between trees and ultrametric spaces.

\section{Hypergraph clustering: examples}

\noindent{\bf Hypergraphs.}\quad In the present section we recall the definition of a hypergraph, and consider the two simplest examples of the hypergraph clustering procedure.

\medskip

A hypergraph is a set $\Gamma$
with a selected system of finite sets $E$ consisting of subsets
containing two or more elements of $\Gamma$.
The elements of $\Gamma$ are called hypergraph vertices,
the sets in $E$ are called hypergraph edges.

If all the edges in $E$ are of cardinality two,
then the hypergraph is a graph.

\medskip

The direct product of two graphs $(\Gamma_1,E_1)$ and  $(\Gamma_2,E_2)$ is a hypergraph with the set of vertices $\Gamma_1\times\Gamma_2$
and with the edges of orders 2 and 4 of the following forms. Let the first and second graphs contain the
respective edges $(A_1,B_1)$ and  $(A_2,B_2)$. With this pair of edges, we associate four 2-edges of the product
hypergraph that are the rows and columns of the $2\times 2$ matrix
$$
\begin{array}{|c|c|}\hline
A_1\times A_2 & A_1\times B_2 \cr\hline
B_1\times A_2 & B_1\times B_2\cr\hline
\end{array}.
$$
The set of all entries of this matrix is a 4-edge. We define the set of the product hypergraph edges using
this procedure: the 2-edges are products of the vertices of one graph by the edges of the other graph, and
the 4-edges are products of the edges of the multiplied graphs.

In general, the direct product of the two hypergraphs $(\Gamma_1,E_1)$ and  $(\Gamma_2,E_2)$ is a hypergraph with the set of vertices $\Gamma_1\times\Gamma_2$ and the set of edges
$$
\Gamma_1\times E_2\bigcup E_1\times\Gamma_2\bigcup E_1\times E_2.
$$

\medskip

\noindent{\bf Hypergraph clustering.}\quad Before the introduction of a general definition of hypergraph clustering we consider several examples.
The general idea of our approach is that the higher order edges are related to cycles in the union of the clustering trees. These cycles describe the different histories of growth of the cluster generated by an increase of the diameter of this cluster with respect to the different metrics.

\bigskip

\noindent{\bf Example 1.}\quad Let us consider the case of a set of three points $A$, $B$, $C$ in the two--dimensional
real plane $\mathbb{R}^2$ with the standard metric. The parameters defining the metric are the
coordinates of the points in the plane.

\medskip

Assume that the set of clusters (vertices of the cluster tree) contains the clusters
$A$, $B$, $C$, $AB$, $ABC$ \footnote{where we denote by $ABC$ the cluster containing $A$, $B$ and $C$}, and the edges of the tree join the vertices in
accordance with the growth of the clusters -- the cluster tree contains the edges
$$
(A, AB),\quad (B, AB),\quad (AB, ABC),\quad (C, ABC).
$$
This defines the tree ${\cal A}_1$ of clusters.

\medskip

Let us consider the variation of the metric (motion of the points in the plane $\mathbb{R}^2$), which replaces the above cluster set with the set of clusters $A$, $B$, $C$, $AC$, $ABC$ with the corresponding edges
$$
(A, AC),\quad (C, AC),\quad (AC, ABC),\quad (B, ABC).
$$
This defines the tree ${\cal B}_1$ of clusters.

\medskip

We define the multidimensional (or hypergraph) clustering in the following way. The set of vertices and 2-edges of the hypergraph ${\cal C}_1$ under discussion is given by the union of the trees of clusters ${\cal A}_1$ and ${\cal B}_1$ defined above (where we identify the clusters which coincide as sets). Namely the vertex set of the hypergraph ${\cal C}_1$ contains the clusters
$$
A,\quad B,\quad C,\quad AB,\quad AC,\quad ABC
$$
and the set of 2-edges (two-point edges) of ${\cal C}_1$ has the form
$$
(A, AB), (B, AB), (AB, ABC), (C, ABC), (A, AC), (C, AC), (AC, ABC), (B, ABC).
$$

The hypergraph ${\cal C}_1$ also contains the 3-edges
$$
(B, AB, ABC),\quad (C, AC, ABC)
$$
and the 4-edge ($A$, $AC$, $AB$, $ABC$).

The partial order of vertices is given by the inclusion of clusters. This finishes the definition of ${\cal C}_1$. The set of the points $A$, $B$, $C$ can be called the border of the hypergraph ${\cal C}_1$ (it is the border of both trees ${\cal A}_1$ and ${\cal B}_1$ of clusters).

\medskip

Schematically (see also the next example) the structure of ${\cal C}_1$ is described by the table
$$
\begin{array}{|c|c|c|}\hline
A & AC & C \cr\hline
AB & ABC & \cr\hline
B &  &  \cr
\hline
\end{array},
$$
where the matrix elements are vertices of ${\cal C}_1$, 2-edges connect all the neighbor vertices in the table and the pairs  $(C, ABC)$, $(B, ABC)$.

The edges of the hypergraph ${\cal C}_1$ describe the growth of clusters starting from some vertex. The higher-order edges correspond to cycles in
the graph that is the union of the clustering trees ${\cal A}_1$ and ${\cal B}_1$.

Namely the 4-edge ($A$, $AC$, $AB$, $ABC$) describes the following situation. If we start from the vertex $A$ we can form the two clusters $AB$ and $AC$ in the trees ${\cal A}_1$ and ${\cal B}_1$ correspondingly which contain $A$. These clusters are related to clusterings with respect to the two different metrics. Then, the cluster in ${\cal A}_1$ which contains the cluster $AB$ is the cluster $ABC$, and the cluster in ${\cal B}_1$ which contains $BC$ is again the cluster $ABC$.

\bigskip

\noindent{\bf Example 2.}\quad Let us consider the case of a set of four points $A$, $B$, $C$, $D$ which are located
in the plane $\mathbb{R}^2$ at the vertices of some quadrangle. In this quadrangle, using the clustering with respect to
the plane metric, we select the clusters
\begin{equation}\label{clusters}
A,\quad B,\quad C,\quad D,\quad AB,\quad CD,\quad ABCD.
\end{equation}
The set of 2-edges contains the edges
\begin{equation}\label{edges}
(A, AB), (B, AB), (C, CD), (D, CD), (AB, ABCD), (CD, ABCD).
\end{equation}
This defines the tree ${\cal A}_2$ of clusters.

\medskip

Let us consider a deformation of the mentioned quadrangle (for example, dilation in some direction in the plane
$\mathbb{R}^2$) under which the metric will be transformed to the metric which defines the cluster tree ${\cal B}_2$ which contains the vertices
\begin{equation}\label{clusters2}
A,\quad B,\quad C,\quad D,\quad AC,\quad BD,\quad ABCD
\end{equation}
and the 2-edges
\begin{equation}\label{edges2}
(A, AC), (C, AC), (B, BD), (D, BD), (AC, ABCD), (BD, ABCD).
\end{equation}

Using the trees ${\cal A}_2$ and ${\cal B}_2$ of clusters, we construct the hypergraph ${\cal C}_2$ which contains the unions of the vertex sets and the 2-edges sets in the described trees and also the four 4-edges
$$
(A, AB, AC, ABCD), (B, AB, BD, ABCD),(C, AC, CD, ABCD), (D, BD, CD, ABCD).
$$

Such a hypergraph can be represented schematically by the table
$$
\begin{array}{|c|c|c|}\hline
A & AC & C \cr\hline
AB & ABCD & CD\cr\hline
B & BD & D \cr
\hline
\end{array}.
$$
The matrix entries are the hypergraph vertices, the 2-edges join the neighboring vertices (in the
horizontal and vertical directions), and the 4-edges correspond to the small $2\times 2$ squares containing the
matrix corners and the cluster $ABCD$.

As in the previous example, the 4-edges describe the histories of the growth of one-point clusters with respect to the different clustering trees.

\medskip


\noindent{\bf Product structure in the hypergraph clustering.}\quad Let us show that the hyper\-graph ${\cal C}_2$ described in Example 2 above can be put in the form of the product of two trees of clusters. This product structure reflects the intrinsic multidimensional structure of the data.

Let us consider the two trees ${\cal T}_1$ and ${\cal T}_2$ which are the trees of clusters in the different spaces.
The tree ${\cal T}_1$ contains the vertices (clusters) $x_1$, $y_1$, $x_1y_1$ and the edges ($x_1$, $x_1y_1$), ($y_1$, $x_1y_1$).
The tree ${\cal T}_2$ contains the vertices $x_2$, $y_2$, $x_2y_2$ and the edges ($x_2$, $x_2y_2$), ($y_2$, $x_2y_2$).

Let us put the hypergraph ${\cal C}_2$ in the form of the product of the trees ${\cal T}_1\times {\cal T}_2$.
The vertices $A$, $B$, $C$, $D$ of ${\cal C}$ in this representation will take the form of products of vertices in ${\cal T}_1$, ${\cal T}_2$
$$
\begin{array}{|c|c|}\hline
A & C\cr\hline
B & D\cr\hline
\end{array}=\begin{array}{|c|c|}\hline
x_1\times x_2 & x_1\times y_2 \cr\hline
y_1\times x_2 & y_1\times y_2\cr\hline
\end{array}.
$$

The other vertices (clusters) of the hypergraph ${\cal C}_2$  are unions of the above vertices, for example,
$$AB=\{x_1\times x_2,y_1\times x_2\}=\{x_1, y_1\}\times x_2,$$
$$AC=\{x_1\times x_2,x_1\times y_2\}=x_1\times \{x_2,y_2\}.$$

Here we use the notation $AB=\{A,B\}$ for the cluster which is the union of vertices $A$ and $B$ (we recall that the notation $(\cdot,\cdot)$ is used for edges).

The 2-edges of the hypergraph ${\cal C}_2$ correspond to edges of one of the trees ${\cal T}_1$, ${\cal T}_2$ multiplied by vertices of the
other tree. For example, the edge ($A$, $AB$) is
$$
(x_1\times x_2, \{x_1\times x_2,y_1\times x_2\})=(x_1,x_1y_1)\times x_2.
$$

The 4-edges of the hypergraph are the products of 2-edges of ${\cal T}_1$, ${\cal T}_2$. In particular
$$
\begin{array}{|c|c|}\hline
A & AC\cr\hline
AB & ABCD\cr\hline
\end{array}=$$
$$
=\begin{array}{|c|c|}\hline
x_1\times x_2 & \{x_1\times x_2,x_1\times y_2\}\cr\hline
\{x_1\times x_2,y_1\times x_2\} & \{x_1\times x_2, y_1\times x_2, x_1\times y_2, y_1\times y_2\}\cr\hline
\end{array}=
$$
$$
=\begin{array}{|c|}\hline
x_1\cr\hline
x_1y_1\cr\hline
\end{array}\times \begin{array}{|c|c|}\hline
x_2 & x_2y_2\cr\hline\end{array}.
$$

The representation of the hypergraph ${\cal C}_2$ by the table can be given in the form of the product of the corresponding representations for trees ${\cal T}_1$, ${\cal T}_2$
$$
\begin{array}{|c|c|c|}\hline
A & AC & C \cr\hline
AB & ABCD & CD\cr\hline
B & BD & D \cr
\hline
\end{array}=
$$
$$
={\small \begin{array}{|c|c|c|}\hline
x_1\times x_2 & \{x_1\times x_2,x_1\times y_2\} & x_1\times y_2\cr\hline
\{x_1\times x_2,y_1\times x_2\} & \{x_1\times x_2, y_1\times x_2, x_1\times y_2, y_1\times y_2\} & \{x_1\times y_2,y_1\times y_2\}\cr\hline
y_1\times x_2 & \{y_1\times x_2,y_1\times y_2\}& y_1\times y_2\cr\hline
\end{array}}=
$$
$$
=\begin{array}{|c|}\hline
x_1\cr\hline
x_1y_1\cr\hline
y_1 \cr\hline
\end{array}\times \begin{array}{|c|c|c|}\hline
x_2 & x_2y_2 & y_2\cr\hline\end{array}.
$$
This representation reflects the intrinsic two-dimensional structure of the hypergraph ${\cal C}_2$.

\section{$p$-Adic case}

\noindent{\bf Multidimensional $p$-adic metric.}\quad One of the main examples of hypergraphs of clusters is related to the geometry of balls in multidimensional $p$-adic spaces.
The standard multidimensional ultrametric in $\mathbb{Q}_p^d$ has the form
$$
d(x,y)={\rm max}_{i=1,\dots,d}(|x_i-y_i|_p),\qquad x=(x_1,\dots,x_d), y=(y_1,\dots,y_d).
$$
In paper \cite{quincunx} the following multidimensional deformed metric in $\mathbb{Q}_p^d$ was considered
\begin{equation}\label{deformed_metric}
d_{q_1,\dots,q_d}(x,y)={\rm max}_{i=1,\dots,d}(q_i|x_i-y_i|_p),\qquad p^{-1}<q_i\le 1.
\end{equation}

The unit ball with respect to the metric $d(\cdot,\cdot)$
$$
\mathbb{Z}_p^d=\{x\in\mathbb{Z}_p^d:~|x_i|_p\le 1,~ x=(x_1,\dots,x_d)  \}
$$
and the dilations $p^k\mathbb{Z}_p^d$, $k\in\mathbb{Z}$ of this ball are balls with respect to all ultrametrics (\ref{deformed_metric}) (for all possible choices of the parameters $q_i$). Therefore we can apply the approach of the previous section and consider the hypergraph of clusters (balls) in  $\mathbb{Q}_p^d$ with respect to some family of metrics of the form (\ref{deformed_metric}).

Let us describe the tree of balls for metric (\ref{deformed_metric}).
Assume that for the metric $d_{q_1,\dots,q_d}$ the parameters satisfy the condition $p^{-1}<q_1<\dots<q_d\le 1$. Then the set of all intermediate $d_{q_1,\dots,q_d}$--balls between
$p\mathbb{Z}_p^d$ and $\mathbb{Z}_p^d$ is given by the sequence of balls
\begin{equation}\label{balls}
B_a=\mathbb{Z}_p\times\dots\times\mathbb{Z}_p\times p\mathbb{Z}_p\times\dots\times p \mathbb{Z}_p,
\end{equation}
with $a$ components $\mathbb{Z}_p$ and $d-a$ components $p\mathbb{Z}_p$, $a=0,\dots,d$.

This sequence of balls is related to a complete flag over the field $\mathbb{F}_p$ with $p$ elements, where we consider the natural correspondence between the $a$-dimensional spaces over $\mathbb{F}_p$ and $B_{a}/p\mathbb{Z}_p^d$.

Recall that a flag is an increasing sequence of subspaces of a finite--dimensional vector space. A flag in the space of dimension $d$ is complete if it contains spaces of all dimensions $0,1,\dots,d$.

Analogously, if we consider the metric $d_{q_1,\dots,q_d}$ where some of the parameters $q_i$ coincide, we obtain a sequence of balls between
$p\mathbb{Z}_p^d$ and $\mathbb{Z}_p^d$ related to an incomplete (partial) flag over  $\mathbb{F}_p$.

We consider also a generalization of the metric (\ref{deformed_metric}), given by
\begin{equation}\label{deformed_metric1}
s(x,y)=d_{q_1,\dots,q_d}(Ax,Ay),
\end{equation}
where $d_{q_1,\dots,q_d}$ is given by (\ref{deformed_metric}) and $A$ is a matrix with matrix elements in $\mathbb{Z}_p$ and $|{\rm det}\, A|_p=1$ (i.e. a matrix of linear isometry with respect to the metric $d=d_{1,1,\dots,1}$).

For a metric from the family (\ref{deformed_metric1}) the sequence of balls between $p\mathbb{Z}_p^d$ and $\mathbb{Z}_p^d$ (obtained by a linear transformation of (\ref{balls}))  will be related to an arbitrary flag over the finite field $\mathbb{F}_p$. The set of all balls for the metric (\ref{deformed_metric1}) will be given by translations and dilations by degrees of $p$ of the described sequence of balls between $p\mathbb{Z}_p^d$ and $\mathbb{Z}_p^d$.

\medskip

\noindent{\bf Hypergraph of balls.}\quad Let us fix some family ${\bf s}$ of ultrametrics  of the above form and consider the hypergraph ${\cal C}(\mathbb{Q}_p^d,{\bf s})$, where the vertices are balls (with respect to some of the ultra\-metrics $s\in {\bf s}$), 2-edges connect the two $s$--balls (with respect to the same metric $s$) which are embedded without intermediate $s$--balls. Since $p^k\mathbb{Z}_p^d$, $k\in\mathbb{Z}$ are balls with respect to all the ultrametrics described above, one can take the union of the trees ${\cal T}(\mathbb{Q}_p^d,s)$ of $s$--balls in $\mathbb{Q}_p^d$ for different $s\in {\bf s}$, where we identify the vertices in the different ${\cal T}(\mathbb{Q}_p^d,s)$ ($s$--balls for the different $s$) which coincide as sets. This gives the sets of vertices and 2-edges of ${\cal C}(\mathbb{Q}_p^d,{\bf s})$. The set of vertices possesses the natural partial order given by inclusion of balls.

Let the family ${\bf s}$ of metrics be sufficiently large, say it will contain the metrics $s_i$ with the parameters $p^{-1}<q_{i_1}<\dots<q_{i_d}\le 1$, where for fixed $i$ the indexes $\{i_j\}$, $j=1,\dots,d$ constitute a permutation of $\{1,\dots,d\}$, and the family ${\bf s}$ contains the metrics corresponding to all possible permutations of $\{1,\dots,d\}$.

Hyperedges (edges of higher order) of ${\cal C}(\mathbb{Q}_p^d,{\bf s})$ are constructed as follows. One of the hyperedges in ${\cal C}(\mathbb{Q}_p^d,{\bf s})$, which we denote by ${\cal D}_d$, is given by the union (for all $s\in{\bf s}$) of the sets of $s$--balls lying between $p\mathbb{Z}_p^d$ and $\mathbb{Z}_p^d$ (including $p\mathbb{Z}_p^d$ and $\mathbb{Z}_p^d$). This hyperedge possesses the structure of a partially ordered graph described above.

Smaller hyperedges ${\cal E}\subset {\cal D}_d$ can be introduced as follows. Let us fix a subfamily ${\bf r}\subset {\bf s}$ of ultrametrics on $\mathbb{Q}_p^d$. Let us fix some ${\bf r}$--ball  $I\in{\cal D}_d$ (i.e. $I$ is an $s$--ball with respect to all $s\in{\bf r}$), which is strictly less than $\mathbb{Z}_p^d$ (in particular, $I\supset p\mathbb{Z}_p^d$). Let $J$ be a smallest ${\bf r}$--ball in ${\cal D}_d$ which is strictly greater than $I$ (since $\mathbb{Z}_p^d$ is an ${\bf r}$--ball, the ball $J$ does exist, the uniqueness of $J$ follows from the ultrametricity of $s\in{\bf r}$). We define ${\cal E}$ as a family $\{K:I\subset K\subset J\}$ of $s$--balls for $s\in{\bf r}$ (i.e. any $K$ is an $s$--ball for some $s\in{\bf r}$). In particular, $I={\rm min}({\cal E})$, $J={\rm max}({\cal E})$.

Other hyperedges in ${\cal C}(\mathbb{Q}_p^d,{\bf s})$ are given by translations and dilations of the hyperedges ${\cal E}$ considered as described above finite sets of balls in $\mathbb{Q}_p^d$.

\medskip

\noindent{\bf Compatible families of ultrametrics.}\quad  We say that the family ${\bf s}$ of ultrametrics on $\mathbb{Q}_p^d$ is compatible, if for any two balls, an $s$--ball $I$ and an $r$--ball $J$, $s,r\in{\bf s}$, the intersection $I\bigcap  J$ is a ball with respect to some ultrametric $t\in{\bf s}$.

The property of compatibility is not satisfied automatically for an arbitrary family ${\bf r}$ of ultrametrics. As we discussed above, ultrametrics on $\mathbb{Q}_p^d$ are related to flags over the finite field $\mathbb{F}_p$. For a family ${\bf r}$ of flags the intersection of some spaces from the different flags in ${\bf r}$ might not be a space from some flag in ${\bf r}$.

\medskip

\noindent{\bf Embedding of hypergraphs of clusters into $p$-adic hypergraphs of balls.}\quad Let us show that the hypergraphs discussed in the previous section can be embedded into a hypergraph associated with a family of multidimensional $p$-adic metrics. Let us consider the quadruple of points in $\mathbb{Q}_2^2$
$$
\begin{array}{|c|c|}\hline
A & C\cr\hline
B & D\cr\hline
\end{array}=\begin{array}{|c|c|}\hline
(0,0) & (0,1) \cr\hline
(1,0) & (1,1)\cr\hline
\end{array}
$$
and perform the clustering procedure with respect to the pair of metrics in $\mathbb{Q}_2^2$ of the form $d_{1,q}(\cdot,\cdot)$, $d_{q,1}(\cdot,\cdot)$, $1/2< q < 1$.

It is easy to see that the metric $d_{q,1}$ will generate the tree ${\cal A}_2$ of clusters with the set of clusters (\ref{clusters}) and the set of edges (\ref{edges}), analogously, the metric $d_{1,q}$ will generate the tree ${\cal B}_2$ of clusters with the set of clusters (\ref{clusters2}) and the set of edges (\ref{edges2}).

Therefore clustering with respect to this pair of metrics generates the hypergraph ${\cal C}_2$ described in Example 2 in the previous section. The product structure of the hypergraph ${\cal C}_2$ described above obtains in this way the natural interpretation of a 2-dimensional structure of $\mathbb{Q}_2^2$.

The hypergraph ${\cal C}_2$ possesses the natural embedding into the hypergraph of clusters in  $\mathbb{Q}_2^2$ with respect to the pair of metrics $d_{q,1}$, $d_{1,q}$. The correspondence between the minimal vertices in ${\cal C}_2$ and balls in  $\mathbb{Q}_2^2$ is given by
$$
\begin{array}{|c|c|}\hline
A & C\cr\hline
B & D\cr\hline
\end{array}\mapsto \begin{array}{|c|c|}\hline
(0,0) & (0,1) \cr\hline
(1,0) & (1,1)\cr\hline
\end{array}+2\mathbb{Z}_2^2=\begin{array}{|c|c|}\hline
(2\mathbb{Z}_2,2\mathbb{Z}_2) & (2\mathbb{Z}_2,1+2\mathbb{Z}_2) \cr\hline
(1+2\mathbb{Z}_2,2\mathbb{Z}_2) & (1+2\mathbb{Z}_2,1+2\mathbb{Z}_2)\cr\hline
\end{array}.
$$
The balls correspondent to non--minimal vertices in ${\cal C}_2$ are constructed as the corresponding unions of the above balls.

Analogously, if we restrict the  hypergraph clustering procedure related to the pair of metrics $d_{q,1}$, $d_{1,q}$ to the set of the three points
$$
\begin{array}{|c|c|}\hline
A & C\cr\hline
B & \cr\hline
\end{array}=\begin{array}{|c|c|}\hline
(0,0) & (0,1) \cr\hline
(1,0) &  \cr\hline
\end{array},
$$
we will get the hypergraph ${\cal C}_1$ described in the Example 1.

\section{Hypergraph of balls for general ultrametric spaces}

\noindent{\bf Hypergraph of balls.}\quad
In the present section we generalize the approach of the previous section to the case of general locally compact ultrametric spaces.

Let $X$ be a locally compact ultrametric space with some family of ultrametrics ${\bf s}$ defined on $X$. Moreover, let, for any pair of metrics $s,r\in{\bf s}$, any $s$--ball be a finite union of $r$--balls. In particular all metrics in ${\bf s}$ define the same topology on $X$. We call $X$ a multidimensional ultrametric space. An example of a multidimensional ultrametric space is given by the space $\mathbb{Q}_p^d$ with the family (\ref{deformed_metric}) of metrics considered in the previous section.

We define the partially ordered hypergraph ${\cal C}(X,{\bf s})$ in a way similar to the one we used for the $p$-adic case. The hypergraph ${\cal C}(X,{\bf s})$ as a graph is a union of the trees ${\cal T}(X,s)$ of $s$--balls, $s\in{\bf s}$. Namely the set of vertices of ${\cal C}(X,{\bf s})$ is the union of the sets of $s$--balls, $s\in{\bf s}$, edges connect $s$--balls (with the same $s$) nested without intermediates. The partial order is by the inclusion of subsets in $X$. If some $s$--ball coincides with some $r$--ball as a set, they define the same vertex in ${\cal C}(X,{\bf s})$.

The family ${\bf s}$ of ultrametrics on $X$ is compatible, if for any two balls, an $s$--ball $I$ and an $r$--ball $J$, $s,r\in{\bf s}$, the intersection $I\bigcap  J$ is a ball with respect to some ultrametric $t\in{\bf s}$.

Hyperedges ${\cal E}$ in ${\cal C}(X,{\bf s})$ are introduced as follows. Let us fix a subfamily ${\bf r}\subset {\bf s}$ of ultrametrics on $X$. Let us fix some ${\bf r}$--ball  $I$ (i.e. $I$ is an $s$--ball with respect to all $s\in{\bf r}$). Let $J$ be a smallest ${\bf r}$--ball which is strictly greater than $I$. We define ${\cal E}$ as a family $\{K:I\subset K\subset J\}$ of $s$--balls for $s\in{\bf r}$ (i.e. any of $K$ is an $s$--ball for some $s\in{\bf r}$). In particular $I,J\in {\cal E}$.

Let us note that for an ${\bf r}$--ball $I$ the minimal ${\bf r}$--ball $J$, $J\supset I$ does not necessarily exist (such a ball always exists for ultrametric spaces containing a finite number of points, if $I$ does not coincide with the whole space). If such an ${\bf r}$--ball $J$ exists, it is uniquely defined.

The introduced hyperedges possess the natural partial order by the inclusion of sets of balls.

\medskip

\noindent{\bf Dimension of an hyperedge.}\quad For the $p$-adic hypergraphs of balls considered in the previous section we have a natural definition of dimension. In this case the dimension of a hyperedge is the number of $p$-adic parameters which one can use for the description of this hyperedge. Let us discuss a notion of dimension which is applicable for general hypergraphs of balls in multidimensional ultrametric spaces.

Let $X$ be a multidimensional ultrametric space with a family ${\bf s}$ of ultrametrics.
Let us consider an ${\bf r}$--hyperedge ${\cal E}\in{\cal C}(X,{\bf s})$, ${\bf r}\subset {\bf s}$, with the minimal ${\bf r}$--ball $I$ and the maximal ${\bf r}$--ball $J$.
There are two properties of $p$-adic hyperedges which one can generalize for the general case:

A) The length of a maximal sequence of nested balls between the minimal and the maximal balls in an hyperedge;

B) The number of maximal subballs in $J$ (with respect to metrics $r\in {\bf r}$) which contain the ball $I$. Here the different maximal subballs in $J$ will be balls with respect to the different metrics $r\in {\bf r}$.

This observation implies the following definition.

\begin{definition}  Let ${\cal E}$ be a ${\bf r}$--hyperedge in ${\cal C}(X,{\bf s})$, ${\bf r}\subset {\bf s}$, with the minimal ${\bf r}$--ball $I$ and the maximal ${\bf r}$--ball $J$.

The A--dimension of the hyperedge ${\cal E}$ is the maximum of the lengths of increasing paths in ${\cal E}$ from $I$ to $J$ (with respect to the partial order in ${\cal E}$) \footnote{The length of a path in a graph is the number of edges in this path.}.

The B--dimension of the hyperedge ${\cal E}$ is the number of balls $J_{k}$, where $I\subset J_k\subset J$ and $J_k$ is a maximal subball of $J$ with respect to some metric $r\in{\bf r}$.
\end{definition}

\noindent{\bf Example.}\quad Let us consider the space $\mathbb{Q}_p^d$ with the family ${\bf s}$ of metrics (\ref{deformed_metric}), which is sufficiently large in the sense described in section 3. In this case we have the maximal (with respect to the partial order on hyperedges) ${\bf s}$--hyperedge ${\cal D}_d$ with the minimal ${\bf s}$--ball $p\mathbb{Z}_p^d$ and the maximal ${\bf s}$--ball $\mathbb{Z}_p^d$.

Both A--dimension and B--dimension of this hyperedge will be equal to $d$. Therefore these dimensions will coincide with the number of $p$-adic coordinates in $\mathbb{Q}_p^d$.

\medskip

For a hypergraph of balls related to a general  multidimensional ultrametric space $X$ with a family of metrics ${\bf s}$, different maximal hyperedges may have different dimensions, and it is possible that the A--dimension and the B--dimension of a hyperedge may be different.

\medskip

\noindent{\bf Embeddings of hypergraphs of balls.}\quad
Let $X$ be a (locally compact) multidimensional ultrametric space with a family ${\bf s}$ of ultrametrics. Let the same conditions hold for the space $Y$ and the family ${\bf r}$ of ultrametrics. We consider the corresponding hypergraphs ${\cal C}(X,{\bf s})$, ${\cal C}(Y,{\bf r})$ of balls.

We assume that there exists a one to one correspondence between the set ${\bf s}$ of ultrametrics on $X$ and some subset of the set ${\bf r}$ of ultrametrics on $Y$. With this one to one correspondence we will use the notation ${\bf s}\subset {\bf r}$. We consider the embedding of the above multidimensional ultrametric spaces as the injective map $i:X\to Y$, for which any $s$--ball $I$ in $X$ maps to a subset of an $s$--ball $J$ in $Y$ with the same diameter and moreover the diameters of $I$ and the image of $I$ in $J$ coincide. The embedding defined in this way is an ${\bf s}$--isometry, i.e. an $s$--isometry with respect to all $s\in {\bf s}$.

At the end of the previous section we have discussed the example of embedding of multidimensional ultrametric spaces and the corresponding embedding of trees and hypergraphs of balls. In a general case, it might happen that the corresponding map at the level of trees and hypergraphs of balls does not exist.
Let us consider the embedding $i:X\to Y$ of ultrametric spaces and let ${\cal T}(X,s)$, ${\cal T}(Y,s)$ be the corresponding trees of balls. Let $I$, $J$, $I\subset J$ be a pair of balls in $X$ nested without intermediates (i.e. the corresponding vertices in ${\cal T}(X,s)$ are connected by edge). Then it is possible that the images $i(I)$ and $i(J)$ are nested with intermediates, i.e. there exists a ball $K\in {\cal T}(Y,s)$: $i(I)\subset K\subset i(J)$. In this case the edge $IJ$ can not map onto an edge in ${\cal T}(Y,s)$.

\medskip

We say that an ultrametric $r$ on the set $X$ is a small deformation of an ultrametric $s$ on $X$ if these ultrametrics generate the same trees of balls, i.e. we have ${\cal T}(X,r)={\cal T}(X,s)$. The definition of a small deformation of a family ${\bf r}$ of ultrametrics on $X$ is analogous --- a family ${\bf r}$ of ultrametrics is a small deformation of a family ${\bf s}$ of ultrametrics  iff  ${\cal C}(X,{\bf r})={\cal C}(X,{\bf s})$.

The next problem discusses, whether it is possible to consider, up to a small deformation of a family of metrics, a finite multidimensional ultrametric space as a subset of $\mathbb{Q}_p^d$ with the family (\ref{deformed_metric}) of ultrametrics.

\medskip

\noindent{\bf Problem.}\quad Let $X$ be a finite multidimensional ultrametric space (i.e. containing a finite number of points) with a family ${\bf s}$ of ultrametrics.

Is it possible to find a small deformation of ${\bf s}$ such that there exists an embedding of $X$ into the multidimensional ultrametric space $\mathbb{Q}_p^d$ for some $p$, $d$, and a family of ultrametrics of the form (\ref{deformed_metric})?

\medskip

Let us note here that we do not claim that the hypergraph ${\cal C}(X,{\bf s})$ can be embedded to the corresponding hypergraph of balls in $\mathbb{Q}_p^d$.
If for a space $(X,{\bf s})$ the above problem possesses a positive solution, we say that the multidimensional ultrametric space $(X,{\bf s})$ is embeddable.

\section{Discussion}

Given a set $X$ with a family of metrics defined on this set one can construct the corresponding trees of clusters and ultrametric spaces described by these trees. When the set $X$ is finite (this condition is satisfied in applications to data analysis) the corresponding ultrametric spaces will possess a natural one to one correspondence with $X$. We obtain a multidimensional ultrametric space $X$ with a family of ultrametrics ${\bf s}$, and the corresponding set of cluster trees ${\cal T}(X,s)$, $s\in {\bf s}$.

Then we can apply to the collection ${\cal T}(X,s)$ the analysis described in the present paper and construct the hypergraph ${\cal C}(X,{\bf s})$ of clusters. This hypergraph is a directed acyclic graph (a graph with a partial order without directed cycles), the (non directed) cycles describe the different possible histories of growth of a cluster with respect to different metrics in ${\bf s}$. Taking into account all possible subsets of the set of metrics ${\bf s}$, we generalize the construction of cycles in the graph of clusters to the construction of hyperedges in ${\cal C}(X,{\bf s})$.

The set $X$ of data may be generated in a complex way, in particular, there may be some independent contributions. In mathematics independence is described by a
dimensionality. A hypergraph is a multidimensional generalization of a graph (in particular, a product of graphs is a hypergraph).

The idea of the approach of the present paper is that there should be some way to describe independencies in data at the level of graphs (and hypergraphs) of clusters. Classification trees (such as trees of clusters) describe the diversity of data, the multidimensional generalization proposed in the present paper should describe the situation where we have independent sources of diversity. In particular, the dimension of an hyperedge will describe the number of sources of diversity (let us note that one can use both the A--dimension and the B--dimension of hyperedges to discuss this subject).

\medskip

One of the applications of classification trees is in bioinformatics. Clustering procedures are applied in bioinformatics in order to generate phylogenetic trees (a phylogenetic tree is a classification tree which is considered as an inferred evolutionary tree).
The metric for the clustering procedure will be equal to the sum of the contributions from the different genetic markers
\begin{equation}\label{markers}
d(X,Y)=\sum_{j=1}^N w_j d_j(X,Y),
\end{equation}
where $w_j\ge 0$ are weights, $X$ and $Y$ are genomes, $d_j$ measure the distance between the genomes for the $j$-th genetic marker (some subsequence of a genome).

Since one may use the different weights $w_j$ for contributions to the classification metric $d(\cdot,\cdot)$ from the different genetic markers, the tree of clusters generated in this way will be essentially non unique. In particular, taking all weights $w_j$ except a single weight to be equal to zero, we obtain the genetic distance measured for a fixed genetic marker.

It was found that clustering (ar analogous procedures of construction of classification trees) applied to the different parts of genomes (and, in general, clusterings with the different parameters $w_j$) may generate different trees. The non-uniqueness of phylogenetic trees will be important in the situations where some parts of a genome have different origins, e.g.  for the cases of reticulate evolution such as hybridization, endosymbiosis or horizontal gene transfer (when some parts of the genome are transferred from the different species).
It was proposed to use the ''forest of life'' (or ''phylogenetic network'') instead of the ''tree of life''  point of view to describe such kind of phenomena, see \cite{KKW} for a review of general applications of networks in biology and \cite{TreeLife,Koonin} for a discussion of phylogenetic trees and evolution.
For the review of mathematical methods for phylogenetic networks one can mention \cite{Huson} and works by A.Dress and coauthors \cite{Clusters,DressBook}.

\medskip

\noindent{\bf Example.}\quad Let us consider the metric (\ref{markers}) for the case of two genetic markers with the corresponding metrics $d_1(\cdot,\cdot)$ and $d_2(\cdot,\cdot)$ and the total metric $d=w_1d_1+w_2d_2$. Assume that each of the two genetic markers may take two possible values which we denote by 0 and 1 and the corresponding distance between 0 and 1 will be equal to one. We have the four possible variants of a genome (four possible pairs of genetic markers)
$$
A=(0,0),\quad B=(1,0),\quad C=(0,1),\quad D=(1,1).
$$

Then, varying the weights $w_1$ and $w_2$,  we obtain the cluster system (\ref{clusters}), (\ref{edges}), described in the Example 2 of Section 2. This cluster system will have the dimension two which corresponds to the presence of the two genetic markers which can vary independently.

\medskip

One of the problems which arise in the consideration of phylogenetic networks is to construct these networks and to embed trees (obtained,
in particular, by clustering of genetic sequences) into these graphs. In our approach we can generate graphs of clusters with cycles using the introduced hypergraph clustering procedure. The embedding of the corresponding phylogenetic trees (obtained by fixing of one metric from the family of metrics used for clustering) is obtained automatically.

In our approach we combine all the trees from the ''forest of data'' (in particular, ''forest of life'') in a single multidimensional hypergraph structure, a ''hypergraph of life''. Phylogenetic networks describe the diversity of genetic information. The application of the hypergraph clustering allows us to investigate the dimensions of hyperedges of the phylogenetic  hypergraph. These dimensions (A--dimensions and B--dimensions) will describe the number of sources of genetic diversity for the corresponding parts of a genome.

\section{Appendix: Ultrametric spaces and trees}

In this Section we discuss the clustering procedure and some results in ultrametric analysis,
which can be found in particular in \cite{ACHA}. A review of some results of $p$-adic mathematical physics can be found in \cite{review}.

Let us recall the definition of clustering. The clustering procedure generates a partially ordered tree of clusters. In this tree vertices are clusters, partial order is defined by inclusion of clusters, an edge connects two clusters nested  without intermediate clusters. The border of this tree is an ultrametric space with the ultrametric defined by the chain distance.

\begin{definition}
A sequence of points $a=x_0,x_1,\dots,\allowbreak x_{n-1},x_n=b$ in a metric space $(M,\rho)$ is called an $\varepsilon$-chain connecting two points
~$a$ and ~$b$ if $\rho(x_k,x_{k+1})\le\varepsilon$ for all $0\le k <n$, and some $\varepsilon>0$. If there exists an $\varepsilon$-chain connecting
$a$ and~$b$ then~$a$ and~$b$ are $\varepsilon$-connected.
\end{definition}

Let $(M,\rho)$ be an arbitrary metric space.
Then the chain distance $d(a,b)$ between $a$ and $b$ is defined by:
$$d(a,b)=\inf(\varepsilon:~a,~b~ {\rm are}~\varepsilon-{\rm connected}).
$$
This distance has all the properties of an ultrametric except for the non--degeneracy property. In particular it satisfies the strong triangle inequality
\begin{equation}\label{strong}
d(a,b)\le {\rm max} (d(b,c),d(a,c)),\quad \forall a,b,c.
\end{equation}

The cluster $C(i,R)$ in a metric space
$(M,\rho)$ is the ball with the center~$i$
and radius $R$ with respect to the chain distance, i.e. the set $\{j\in M\colon  d(i,j)\le R \}$.

\begin{definition}
The clustering of the space~$M$ is a set of clusters in~$M$ such that:

i) every element in~$M$ belongs to some cluster;

ii) for any pair $a$, $b$ of elements in~$M$ there exists a minimal cluster $\sup(a,b)$ containing both elements;

iii) for arbitrary embedded clusters $A\subset B$  every increasing
sequence of embedded clusters $\{A_i\}$,
$A\subset\dots\subset A_i\subset\dots\subset B$ is finite;

iv) the total number of clusters in the clustering is finite or countable.
\end{definition}

\noindent {\bf Example.}\qquad  Let $D=\{d_i\}$ be a countable set
of positive numbers without positive accumulation points. Consider
the clustering ${\cal C}_D$ of the metric space $(M,\rho)$ which
contains all clusters of chain radii $d_i\in D$ and arbitrary
centers.

\medskip

An ultrametric space is a metric space with the metric $d(x,y)$ satisfying the strong triangle inequality (\ref{strong}).
Ultrametric spaces are dual to trees with some partial order. Below we describe
some part of the duality construction.

For a (complete locally compact) ultrametric space $X$ we consider the set ${\cal T}(X)$, which is the result of clustering of $X$ with respect to the ultrametric, i.e. ${\cal T}(X)$ contains all the balls in $X$ of nonzero diameters, and the
balls of zero diameter which are maximal subbals in balls of nonzero
diameters. This set possesses a natural structure of a partially ordered
tree. The partial order in ${\cal T}(X)$ is defined by inclusion of balls.

Two vertices $I$ and $J$ in ${\cal T}(X)$ are connected by an
edge if the corresponding balls are ordered by inclusion, say
$I\supset J$ (i.e. one of the balls contains the other), and there
are no intermediate balls between $I$ and $J$.

On the tree ${\cal T}(X)$ we have the natural increasing
positive function which associates to any vertex the
diameter of the corresponding ball.

Assume now that we have a partially ordered tree ${\cal T}$,
satisfying the conditions:

1) Graph ${\cal T}$ is a tree, i.e. for any pair of vertices there
exists a finite path in ${\cal T}$ which connects these vertices
and ${\cal T}$ does not contain cycles.

2) Each vertex in ${\cal T}$ is incident to a finite set of edges.

3) For any finite path in ${\cal T}$ there exists a unique maximal vertex in this path.

Let us choose an arbitrary  positive increasing (w.r.t. the
partial order) function $F$ on this tree. Then we define the
ultrametric on the set of vertices of the tree ${\cal T}$ as
follows: $ d(I,J)=F({\rm sup}(I,J)), \,$ (for $I \ne J$),  where
${\rm sup}(I,J)$ is the supremum of vertices $I$, $J$ with respect
to the partial order. The vertex ${\rm sup}(I,J)$ coincides with
the above mentioned unique maximal vertex in the path $IJ$.

Then we take the completion of the set of vertices with respect to the
defined ultrametric and eliminate from the completion all the inner
points of the tree (a vertex of the tree is inner if it does not
belong to the border of the tree, i.e. it is incident to more than one edge). We denote the obtained space by
$X({\cal T})$, this space is ultrametric, complete and locally compact. The space $X({\cal T})$ is called the border of the tree ${\cal T}$.

\bigskip

\noindent{\bf Acknowledgments}\qquad This work is partially supported by the DFG project AL 214/40-1.
One of the authors (S.K.) gratefully
acknowledges being partially supported by the grants of
the Russian Foundation for Basic Research
RFBR 11-01-00828-a and 11-01-12114-ofi-m-2011, by the grant of the President of Russian
Federation for the support of scientific schools NSh-2928.2012.1,  and
by the Program of the Department of Mathematics of the Russian
Academy of Science ''Modern problems of theoretical mathematics''.

\end{document}